\documentstyle[multicol,epsfig,aps,amssymb]{revtex}
\newcommand{\rhobar}{{\overline{\rho}}}
\newcommand{\calN}{{\mathcal{N}}}

\begin{document}

\title{Velocity and diffusion coefficient of $A+A\leftrightarrow A$
reaction fronts in one dimension}

\author{Niraj Kumar and Goutam Tripathy } 
\address{Institute of Physics, Sachivalaya Marg, Bhubaneswar 751005,
India} 

\maketitle

\begin{abstract}

 We study front propagation in the reversible reaction-diffusion system
$A + A \leftrightarrow A$ on a 1-$d$ lattice. Extending the idea of
{\it leading} particle in studying the motion of the front we write a
master equation in the stochastically moving frame attached to this
particle. This approach provides a systematic way to improve on
estimates of front speed obtained earlier. We also find that the
leading particle performs a correlated random walk and this
correlation needs to be taken into account to get correct value of the
front diffusion coefficient.

\end{abstract}

\begin{multicols}{2}

\section{Introduction}
\begin{figure}
\includegraphics[width=9cm]{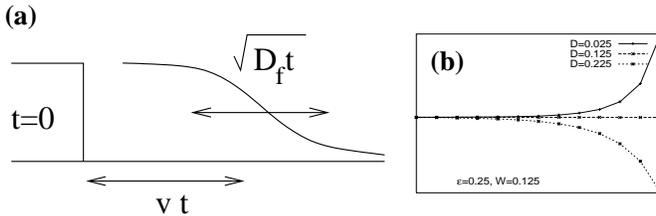}
\caption{Schematic picture of the development of front in the
reaction-diffusion system. (a): The ensemble averaged front propagtes with
a unique asymptotic speed $v$ and spreads diffusively with a
coefficient $D_f$. (b) If the ensemble average is taken in a frame
moving with the leading particle in each realization, the profile
obtained is a time-invariant profile. The three profiles shown for
different values of the parameters of the microscopic model discussed
later in the text.}

\label{fig:front}
\end{figure}

Propagation of fronts into unstable states in reaction-diffusion
systems has been an actively studied topic for a long
time\cite{reviews}. The dynamics of propagating fronts is of interest
in various diverse physical situations \cite{recentworks}.  Recently,
there is renewed activity concerning connection between fronts in
microscopic discrete stochastic models and macroscopic deterministic
equations which are believed to be their mean-field limits
\cite{largeN}. In this paper we study front propagation in a
system of reacting and diffusing particles on an infinite 1-$d$
lattice (Fig.\ref{fig:front}). Initially, left half of the lattice is
filled with a certain density of particles with the right half being
completely empty,  resulting in a sharp boundary
between the two halves. As the system evolves, the particles move to
the right resulting in a propagating front. After an initial
transient, the front moves with an asymptotic speed $v$. Further, due
to the inherent stochastic nature of the dynamics, the ensemble
averaged front profile undergoes diffusive broadening
(\ref{fig:front}a) with an associated diffusion coefficient $D_f$.
There are a number of alternative definitons of the position of the
front in any given realization of the process but it is found that all
of them lead to the same values of $v$ and $D_f$. The simplest choice,
which is also the easiest to implement numerically, is to take the
position of the rightmost particle at any time to be the location of
the front \cite{ker1,ker2,ptvs}.  Thus, the dynamics of the front is
reduced, in an approximate way, to the dynamics of a single particle -
the {\it leading} particle. Treating the motion of this particle as a
biased random walk, simple approximate expressions for the speed $v$
and diffusion coefficient $D_f$ has been derived
\cite{ptvs,ker2}. However, in these approaches it is hard to find a
systematic way to improve estimates for $v$ and $D_f$. In the present
paper we show that by writing the front dynamics in a frame moving
with the leading particle one can get numerically better estimates of
the front speed and diffusion coefficient. We also show that the
motion of the leading particle is correlated in time and thus the
front diffusion coefficient differs from that obtained from a simple
random walk approximation.

\section{Model and results for $v$ and $D_f$}

\begin{figure}
\includegraphics[width=7cm]{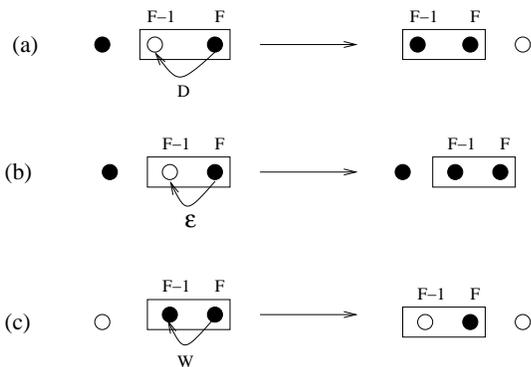}
\caption{Basic microscopic moves of (a) diffusion, (b) birth and (c)
death.  When the right particle happens to be the leading particle
(denoted by F) the processes can result in transitions between the two
states 01 and 11 as described later in the text.}

\label{fig:moves}
\end{figure}

We consider a 1-$d$ lattice ($-\infty < i<\infty$) in which
each site can hold atmost one particle ('hard-core exclusion'). The
particles (denoted by A) undergo three basic microscopic processes:
(i) Birth/creation: a particle can generate a new one on a
neighbouring empty site with rate $\epsilon$, (ii) Death/annihilation:
one of the two neighboring particles gets annhilated with rate $W$,
and (iii) Diffusion: A particle diffuses to a neighbouring empty site
with rate $D$ (see~Fig.\ref{fig:moves}). Initially, at $t=0$, the left
half of the lattice ($i\le 0$) is filled with particles at a density
$\rho=\rhobar$, where $\rhobar=\epsilon/(\epsilon+W)$ is the density
of the equilibrium phase obtained if the process is allowed to occur
in a finite system \cite{footnote1}. There are only two independent
parameters in the system as one of the three rates $\epsilon,D,W$ can
be scaled away by choosing the time scale appropriately. In all our
simulations $\epsilon=0.25$ and the two ratios $D/\epsilon$ and
$W/\epsilon$ are of the order of unity.  As was noted in \cite{ptvs},
this choice is fairly generic for the diffusion limited fluctuation
dominated regime we would like to focus on.

In the mean-field description, which is expected to be valid in the 
limit $D\rightarrow\infty$  and the evolution
of the front is described by the F-KPPP  equation \cite{fkpp}
$\partial_t\rho=D\partial_x^2\rho+\epsilon\rho-(\epsilon+W)\rho^2$,
where $\rho$ is the coarse grained density of A particles.
The front speed is then given by $v_0=2\sqrt{\epsilon D}$.

For the case $W=0$, which arises naturally in the context of turbulent
flame front propagation \cite{ker1}, the existence and uniqueness of
an asymptotic front solution were established rigourously in
\cite{bramson} and it was shown that the 'mean-field' limit is
obtained as $D\rightarrow \infty$.  For finite values of $D$, a 'two
particle' representation was used in \cite{ker2} to get an approximate
value of the front speed which works quite well for $D/\epsilon \sim
O(1)$.

\noindent In \cite{bA}, using the inter-particle distribution
functions an exact solution was obtained for the special case
$W=D$. In this case, it was shown that in each realization of the
front evolution, the leading particle performs a biased random walk
and the particles behind it are distributed exactly in equilibrium
with density $\rho=\rhobar$. The speed and the diffusion coefficient
were found to be $v=\epsilon$ and $D_f=\epsilon+D$ respectively.

\noindent If $W\ne D$, the method used in \cite{bA} does not work and
an exact solution is no longer possible. Also, the two particle
representation of \cite{ker2} does not close if particle annihilation
is introduced, i.e, $W > 0$.  In \cite{ptvs}, an approach based
explicitly on the motion of the leading particle as a biased random walker
was introduced and the following approximate expressions for speed and
diffusion coefficient were obtained.
\begin{equation}
v=\epsilon - \rho_1 (W-D); 2D_f=2D+\epsilon-\rho_1 (W-D),
\label{eq:vD-ptvs}
\end{equation}
where $\rho_1$ is the probability of occupation of the site just
behind the leading particle.  Expression (\ref{eq:vD-ptvs}) above may be
written down by noting that the fronts moves right
with rate $P=\epsilon+D$ (i.e. whenever a particle is created to the
right of the leading particle or the later makes a diffusive move to
the right). The leading particle takes  a negative step when it gets
annihilated by the particle on the left (with rate $W\rho_1$) or if it
makes a diffusive move to an empty left site (with rate
$D(1-\rho_1)$), where $\rho_1$ is the occupation probability of the
site immediately to the left of the leading particle. For $W=D$, these
results reduce to those obtained exactly in \cite{bA}. In order to get
more accurate values of $v$ and $D_f$ one needs to find better
estimates of $\rho_1$ (the bulk value $\rho_1=\rhobar$ was used in
\cite{ptvs}).  While (\ref{eq:vD-ptvs}) works quite well (with
$\rho_1=\rhobar$) for $W$ close to $D$, there are significant
deviations as one moves away from this special point.

\begin{figure}
\includegraphics[width=8cm]{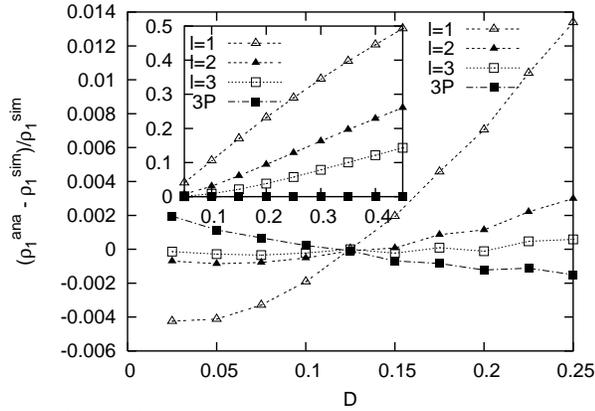}
\caption{Deviation of $\rho_1$ computed via various approximations
from that obtained from direct simulation as a function of $D$ for
$W=0.125$ and $W=0$ (inset). Empty squares correspond to the
3-particle representation discussed in subsection A. }

\label{fig:rho1}
\end{figure}

\begin{figure}
\includegraphics[width=8cm]{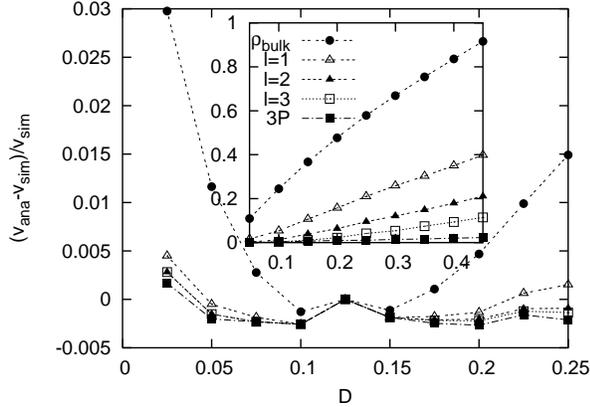}
\caption{Difference between front speed computed using $\rho_1$ from
various approximations and that obtained from direct simulation as a
function of $D$ for $W=0.125$ and $W=0$ (inset). Filled squares
correspond to the 3-particle representation discussed in subsection
A.}

\label{fig:speed}
\end{figure}

\begin{figure}
\includegraphics[width=8cm]{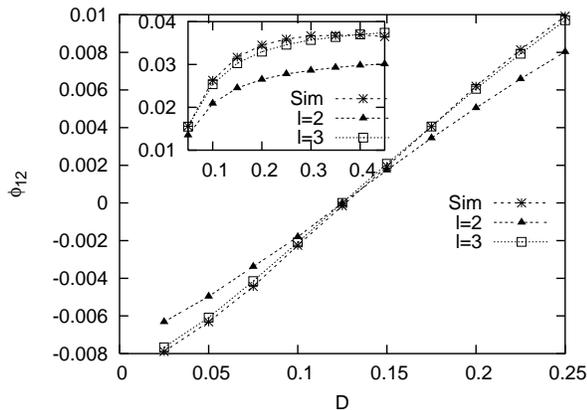}
\caption{The correlation $\phi_{12}$ between occupancies of the pair of
sites immediately following the leading particle as a function of $D$
($W=0.125$). We see that the correlation is negative for $D<W$ and
positive for $D>W$. Inset: $\phi_{12}$ as a function of $D$ for $W=0$.}

\label{fig:dcor}
\end{figure}

\noindent In order to get successively better estimates for $\rho_1$,
we look at the invariant profile of the front as observed from the
leading particle. Starting from an ensemble of $\calN$ realizations
this invariant profile is obtained, after an initial transient time,
by aligning the leading particle of each member of the ensemble
(Fig.~\ref{fig:front}b).  From the definition of $\rho_1$ it then
follows that out of the $\calN$ realizations $\rho_1\calN$ have a
particle in the site to the left of the leading particle and the rest
$(1-\rho_1)\calN$ have an empty site next to it. In the steady state
(which is asymptotically reached after transients) it thus follows
that there is a kinetic balance between the two types of realizations:
those with 11 or 01 as the occupancy of the rightmost pair of sites
(where the second 1 denotes the leading particle). Thus, the two
states 01 and 11 may be thought of as a truncated representation of
the full lattice. Due to the microscopic moves there are transitions
between these two 'states'. For example, in a realization in the 11
state the leading particle makes a diffusive move to the right, the
state of the realization changes to 01 \cite{footnote2}. Considering
all such transitions one can write a master equation for the
probabilities $p_{11}$ and $p_{01}$ in this truncated state space
\cite{foot2a}
\begin{eqnarray}
\dot{p}_{01} & = & (2D-D\rho_b+2W)p_{11}-(2D\rho_b+2\epsilon+\epsilon\rho_b)p_{01}, \nonumber\\
\dot{p}_{11} & = & (2D\rho_b+2\epsilon+\epsilon\rho_b)p_{01}-(2D-D\rho_b+2W)p_{11}.
\label{eq:l=1}
\end{eqnarray}
In the steady  state $\dot{p}_{01}=0=\dot{p}_{11}$ and one obtains
($p_{11}=\rho_1$ in steady state) 
\begin{equation}
 \rho_1=\frac{3\epsilon^2+2\epsilon(W+D)}
 {3\epsilon^2+2W^2+4\epsilon W+3 D \epsilon + 2 D W}
\label{eq:l1rho1}
\end{equation}
where we have used $\rho_b=\rhobar$, an approximation which becomes
better as one includes more sites in the truncated
representation. E.g. we have also computed $\rho_1$ by keeping $l=3$ sites
(i.e. 4 states: 111, 011, 101, 001), $l=4$ sites (8 states) and the values
obtained for $\rho_1$ are plotted in Fig.~\ref{fig:rho1} and the
corresponding front speed (using Eq.(\ref{eq:vD-ptvs})) in
Fig.~\ref{fig:speed}. It is to be noted that expression
(\ref{eq:l1rho1}) above reduces to $\rho_1=\rhobar$ for $W=D$ as it
should.  Further, in this approach it is also possible to obtain the
spatial density correlation between site occupancies $\phi_{12}=\langle n_1 n_2\rangle -\rho_1\rho_2/\rhobar^2$  (Fig. \ref{fig:dcor}).

\subsection{Reduced 3-particle representation}

\noindent The drawback of the above approach is that it is not easy to
write the transition matrix as the number of sites $l$ is increased
(number of states increases as $2^l$). In the following we try to find
an analytically tractable estimate for $\rho_1$ using an alternative
reduced representation similar to Kerstein's approach
\cite{ker2}. Instead of keeping a fixed number of sites to denote a
state, Kerstein chose the following infinite set of two particle
states: \{11, 101, 1001, 10001, $\cdots$ \} where the 'k'th state has
$k$ empty sites between the leading particle (denoted by the 1 on the
right) and the next particle (denoted by the 1 on the left). If there
is no annihilation process, i.e., $W=0$, then these set of states is
closed with respect to transitions between the states. However, as was
pointed out in \cite{ptvs}, this breaks down as soon as $W>0$. To see
this let us consider the 11 state and the microscopic process in which
one of the two particles gets annihilated.  The resulting state
depends upon the location of the third particle in the initial
configuration of the lattice. I.e., one needs to go to a 3-particle
representation. However the same problem occurs while considering
annihilation in the 111 state.  Thus, to take care of this problem
which arises due to the effective non-locality in transition rates by
going to the moving frame attached to the leading particle one needs
to make a reasonable truncation of the hierarchy. We extend the set of
states by including the states \{111, 1011, 10011, $\cdots$ \} where
the rightmost 1 denotes the leading particle. Let us denote by $p_k$
the probability of the two particle states $1\underbrace{0...0}_k 1$
with $k$ empty sites between the leading particle and the next and by
$q_k$ the probability of the three particle state
$1\underbrace{0...0}_k 11$ with $k$ empty sites between the second and
the third particles. The transition rates within these states can be
written down following the procedure as in the case of fixed number of
sites representation discussed above and one obtains the following
rate equations for the $p_k$'s.

\begin{eqnarray}
\dot{p}_k & = & (2D-D\rho+W\rho)p_{k-1}+(2D+\epsilon)p_{k+1} \nonumber\\ 
           & & + (q_{k-1}+q_k)W - (4D-D\rho+3\epsilon+W\rho)p_{k}; 
           ~~~~~k\ge 2, \nonumber \\
\dot{p}_1 & = & (2D-D\rho)p_0+(2D+\epsilon)p_2+(2q_0+q_1)W \nonumber \\
          & & -(4D-D\rho+3\epsilon+W\rho)p_1; \nonumber \\
\dot{p}_0 & = & (2D+\epsilon)p_{1}+2\epsilon(1-p_{0})
  -(2D-D\rho-2W)p_0,
\label{eq:pk}
\end{eqnarray}
\noindent where we made the approximation that $\rho$ is the
probability that the site next to the last particle in the two or
three particle states is occupied independent of its distance from the
leading particle. We do not write the equations for $q_k$'s as it can
be shown that they can be eliminated in the steady state. To obtain
the steady state solution $\dot{p}_k=0$, following Kerstein, we make
the ansatz $p_k = p_0(1-p_0)^k$ and write $\rho=Ap_{0}-Bp_{0}^{2}$ to
take care of the ignored correlations in the ansatz in a
phenomenological way. The parameters $A$ and $B$ are fixed in the
following way. We first note that for $D=W$, both the quantities
$\rho$ and $p_0$ equal the bulk density, namely,
$\rho=p_{0}=\rhobar=\epsilon/(\epsilon+W)$ implying
$A=1+\epsilon B/(\epsilon+W)$ and thus
\begin{equation}
  \rho=(1+\frac{\epsilon B}{\epsilon+W})p_{0}-Bp_{0}^{2}
\label{eq:rho_p0}
\end{equation}
This form also ensures that $W=0$ and $p_{0}$=1 imply $\rho$=1. Using
$\rho$ from Eq.~(\ref{eq:rho_p0}) in Eqs.~(\ref{eq:pk}) (in steady
state $\dot{p}_k=0$) we get the following cubic equation for
$p_{0}$, 
\begin{eqnarray}
  & & DB(\epsilon+W)p_{0}^{3}+(\epsilon^{2}+\epsilon
W+\epsilon D+WD-DB\epsilon)p_{0}^{2} \nonumber\\
 & &  +(\epsilon^{2}+3\epsilon W+2W^{2})p_{0}
  -2\epsilon^{2}-2\epsilon W = 0
\label{eq:cubic}
\end{eqnarray}
\noindent In order to fix $B$ we note that for large $D$
  we expect the front to approach the mean-field limit with the speed
  given by $v_0=2\sqrt{\epsilon D}$. Since, in terms of $p_{0}$ the
  velocity is given by (from eq.~(\ref{eq:vD-ptvs}))
  $v=\epsilon-(W-D)p_{0}\sim Dp_{0}$ for $D>> W,\epsilon$, this
  implies, in this limit,
  $p_{0}=2\sqrt{2\epsilon/D}$.  Substituting this
  expression for $p_{0}$ in Eq.~(\ref{eq:cubic}) we obtain
  $B=3(\epsilon+W)/4\epsilon$.  Using this expression for B
  in Eq.~(\ref{eq:cubic}) one obtains $p_0=\rho_1$
  implicitly through,
\begin{equation}{\label{eq:Dp0}}
  D=\frac{4\epsilon(2\epsilon-\epsilon p_0^2-\epsilon
  p_0-2p_0 W)}{(\epsilon+3\epsilon p_0+3Wp_0)p_0^2}
\end{equation}

The result is shown in Fig.~\ref{fig:rho1} (square symbols
marked as 2P).  We note that, although several ad-hoc assumptions were
made in arriving at eq.~(\ref{eq:Dp0}), including using results that are
strictly valid in the mean-field limit $D\rightarrow\infty$, the
agreement with direct numerical results is remarkable even for the
finite values of $D$ considered.

\subsection{Diffusion coefficient: $D_f$}

\begin{figure}
\includegraphics[width=8cm]{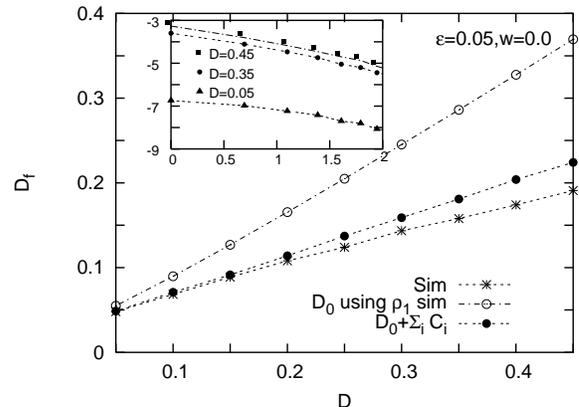}
\caption{Diffusion coefficient of the front for $W=0$ as a function of
$D$.  The bottom data (pluses) are the direct simulation values. The
top (crosses) are the values obtained from Eq.~(\ref{eq:vD-ptvs}) while
the middle one (stars) represent the correlation corrected $D_f$.
Inset: the correlation function $C(t)$ for different values of
$D$.}

\label{fig:diff}
\end{figure}

In Fig.~\ref{fig:diff}, we plot the front diffusion coefficient $D_f$
as a function of $D$ ($W=0$, $\epsilon=0.05$). The lower curve is from
direct simulation and the upper one is that obtained from
Eq.~(\ref{eq:vD-ptvs}) by using the most accurate estimate of
$\rho_1$. We see marked deviation of the values obtained from the
analytic expression which implies that the simple (uncorrelated)
random walk picture of the leading particle in not quite correct. The
motion of a correlated biased random walker is described by
$x(t+1)-x(t)=v+\eta(t)$ where the noise term $\eta$ is temporally
correlated: $\langle\eta(t)\rangle=0$ and
$\langle\eta(t)\eta(t')\rangle\equiv C(t-t')\ne 0$.  The mean speed of
the walker is $v$ and the asymptotic diffusion coefficient is given by
\begin{equation}
D_f=\sum_{t=0}^{\infty} \langle\eta(0)\eta(t)\rangle =
\sum^\infty_{\tau=0} C(\tau).
\label{eq:diff}
\end{equation}
\noindent Indeed, for the front under study we find that there is long range
correlation between the successive steps of the leading
particle (Fig.~\ref{fig:diff}, inset). This correlation is
non-positive for all parameters (both $D>W$ as well as $D<W$) and
vanishes for the special case of $D=W$. Once this correlation is taken
into account the diffusion coefficient matches reasonably with that
obtained from direct simulations for the range of $D$ studied
\cite{footnote3}.  Preliminary fits indicate that the correlation
function has the functional form $C(t)=A~t^{-\alpha}\exp{(-t/\tau)}$.

\section{Conclusion}

 We have illustrated the usefulness of the leading particle
picture in describing the propagation of fronts in the
$A+A\leftrightarrow A$ reaction-diffusion process in the diffusion
controlled limit in one dimension.  By writing the master equation in the
moving frame attached to the leading particle we are able to
obtain better numerical estimates for the density of the site behind
the leading particle and thus the front speed. In addition, this
approach, in principle allows one to compute the spatial density
profile and density-density correlations away from the special point
$D=W$.

\noindent Our numerical results show that the motion of the leading
particle is correlated in time and this needs to be taken into account
in order to get the correct diffusion coefficient. It is seen that
this correction increases with increasing $D$ (the microscopic
particle diffusion constant) and thus it might play an important role in
determining how the mean-field limit is achieved as $D\rightarrow
\infty$.

\end{multicols}


\begin{thebibliography}{02}
\bibitem{footnote1} If any other density is chosen initially it reaches 
$\rhobar$ exponentially fast in time.
\bibitem{footnote2} In this representation the state may not always
change although the configuration of the lattice as a whole
changes. E.g., in the 11 state if a new particle is created to the
right of the leading particle the state still remains as 11.
\bibitem{foot2a} If the move involves the third site (counting to left
from the leading particle), e.g., when the second particle gets
annihilated by the particle on the third site, we assume the third
site is occupied with probability $\rho_b$, independent of whether the
second site is occupied or not.
\bibitem{footnote3} Values of $D_f$ obtained from eq.~(\ref{eq:vD-ptvs})
is essentially $C(0)=\langle \eta^2(0)\rangle$.
\end{thebibliography}
\end{document}